# Doping dependent evolution of magnetism and superconductivity in Eu$_{1-x}$K$_x$Fe$_2$As$_2$ (x = 0-1) and temperature dependence of lower critical field H$_{c1}$

Anupam[1], P.L. Paulose[2], S. Ramakrishnan[2], Z. Hossain[1]

[1]Department of Physics, Indian Institute of Technology Kanpur, Kanpur-208016, India

[2]DCMP&MS, Tata Institute of Fundamental Research, Mumbai-400 005, India

**Abstract**

We have synthesized the polycrystalline samples of Eu$_{1-x}$K$_x$Fe$_2$As$_2$ (x = 0-1) and carried out systematic characterization using x-ray diffraction, ac & dc magnetic susceptibility, and electrical resistivity measurements. We have seen a clear signature of the coexistence of superconducting transition (T$_c$ = 5.5 K) with SDW ordering in our under doped sample viz. x = 0.15. The spin density wave transition observed in EuFe$_2$As$_2$ get completely suppressed at x = 0.3 and superconductivity arises below 20 K. Superconducting transition temperature T$_c$ increases with increase in K content and a maximum T$_c$ = 33 K is reached for x = 0.5, beyond which it decreases again. The doping dependent T(x) phase diagram is extracted from the magnetic and electrical transport data. It is found that magnetic ordering of Eu-moments coexists with superconductivity up to x = 0.6. The isothermal magnetization data taken at 2 K for the doped samples suggest 2+ valence states of Eu ions. We also present the temperature dependence of the lower critical field H$_{c1}$ of superconducting polycrystalline samples. The value of H$_{c1}$(0) obtained for x = 0.3, 0.5, and 0.7 after taking the demagnetization factor into account is 248, 385, and 250 Oe, respectively. The London penetration depth λ(T) calculated from the lower critical field does not show exponential behaviour at low temperature, as would be expected for a fully gapped clean s-wave superconductor. In contrast, it shows a T$^2$ power-law feature down to T = 0.4 T$_c$, as observed in Ba$_{1-x}$K$_x$Fe$_2$As$_2$ and BaFe$_{2-x}$Co$_x$As$_2$.

## Introduction

Interplay between magnetism and superconductivity has been an exciting and very active area of research in condensed matter physics community since last few decades. It was predicted by Ginzburg et al. that long range ferromagnetism (FM) can not persist with superconductivity (SC) while antiferromagnetism (AFM) could coexist with SC [1]. The coexistence of superconductivity and magnetism was observed in ternary and quaternary intermetallic compounds $RMo_6S_8$, $RRh_4B_4$ and $RNi_2B_2C$ (R = rare earth), respectively [2, 3]. Among the rare earth borocarbides, the re-entrant antiferromagnetic superconductor $HoNi_2B_2C$ exhibits an exotic interplay of superconductivity and magnetism. It becomes superconducting at ~ 7.5 K, re-enters the normal conducting state at 5 K, and recovers superconductivity at lower temperature [3]. The interplay between magnetism and superconductivity and the coexistence of two is still a mysterious topic.

The recent discovery of superconductivity (SC) in new Fe-based systems by doping the magnetic parent compound with electrons or holes or by applying pressure and thereby, suppressing the magnetic order, has activated intense research [4]. Among the A-122 series (A= Eu, Sr, Ba, Ca), $EuFe_2As_2$ is an interesting candidate to investigate because it contains two different magnetic atoms. In addition to the Fe sub-lattice, the Eu sub-lattice consists of $Eu^{2+}$ ions with large magnetic moment ($7\mu_B$/fu) [5]. $Eu^{2+}$ spins order ferromagnetically within the ab-plane while A-type antiferromagnetism is observed along the c-axis. The Fe sub-lattice exhibits long-range spin density wave (SDW) ordering accompanied and/or followed by the tetragonal to orthorhombic structural transition below 190 K, whereas the Eu lattice exhibits an antiferromagnetic ordering at $T_N$ = 19 K [5]. Thus $EuFe_2As_2$ presents a good example to study the interaction between superconductivity and magnetic ordering of $Eu^{2+}$ moments. The interplay of pressure-induced superconductivity and the $Eu^{2+}$ order in this compound leads to a re-entrant behaviour similar to as observed in $RNi_2B_2C$ [6]. Superconductivity can be induced in $EuFe_2As_2$ by substitution of K for Eu [7] or by P-doping on As site [8]. The coexistence of $Eu^{2+}$ short range ordering and superconductivity has been observed in $Eu_{0.5}K_{0.5}Fe_2As_2$ as verified by Eu Mossbauer spectroscopy [7]. Also the coexistence of ferromagnetism and superconductivity with a re-entrant type feature has been observed in $EuFe_2As_{1.6}P_{0.4}$ below 20 K [8].

In this paper, we have characterized the well prepared $Eu_{1-x}K_xFe_2As_2$ (x = 0-1) polycrystalline samples through various experimental techniques. We systematically measure the magnetic and transport properties to study the superconductivity and magnetism in the K-doped $EuFe_2As_2$. The results are presented in the form of a T-x phase diagram. The doping dependent phase diagram has been reported in the literature for $A_{1-x}K_xFe_2As_2$ (A=Ba, Sr) and the coexistence of SDW state and structural transition with superconductivity was observed in a narrow temperature range [9, 10]. For the 1111-FeAs systems the coexistence of magnetic ordering and superconductivity has been observed only for R = Sm, while for other rare earths viz. R = La, Pr, Ce the onset of superconductivity appears at a concentration where magnetic transition get suppressed completely [11]. We have also determined the lower critical field $H_{c1}$ and superfluid density for various samples and found that the London penetration depth deviates from the exponential behaviour and follows power law behaviour at low temperatures. A lot of experiments have been carried out to determine the pairing symmetry in FeAs based superconductors but the exact nature of pairing mechanism is not yet settled in any system. The power law behaviour of penetration depth at low temperature has been observed for $Ba_{1-x}K_xFe_2As_2$ [12] and $BaFe_{2-x}Co_xAs_2$ [13] which has been proposed to be due to the unconventional $s^{\pm}$ state. In contrast, the microwave measurements in $Ba_{1-x}K_xFe_2As_2$ suggest a fully gapped state with two gaps [14], whereas µSR measurements show a linear temperature dependence suggesting a nodal gap [15]. Similar contradictory results have been reported for other FeAs based superconducting systems.

**Experimental details**

The polycrystalline samples of $Eu_{1-x}K_xFe_2As_2$ (x = 0-1) were prepared by solid state reaction. High purity elements were taken in stoichiometric ratio inside an alumina crucible which was sealed in a tantalum crucible under argon atmosphere to minimize the loss of potassium by vaporization. The latter was finally sealed inside an evacuated quartz ampoule and subsequently kept under the first heat treatment as described in [7]. The preheated product was grounded thoroughly and pressed into pellets and annealed at 900°C for 5 days. All the sample handling was done inside a glove box filled with argon atmosphere. The samples were characterized by powder x-ray diffraction with Cu $K_\alpha$ radiation to determine the single phase nature and crystal structure. Scanning electron microscope (SEM) equipped with energy dispersive x-ray analysis (EDAX) was used to check the homogeneity and composition of the samples. Low field magnetic measurements were performed in superconducting quantum interference device (SQUID, Quantum design) magnetometer and high field measurements were done using vibrating sample magnetometer (VSM, oxford instruments). Resistivity measurements were carried out using standard four probe technique in a liquid helium cryogen.

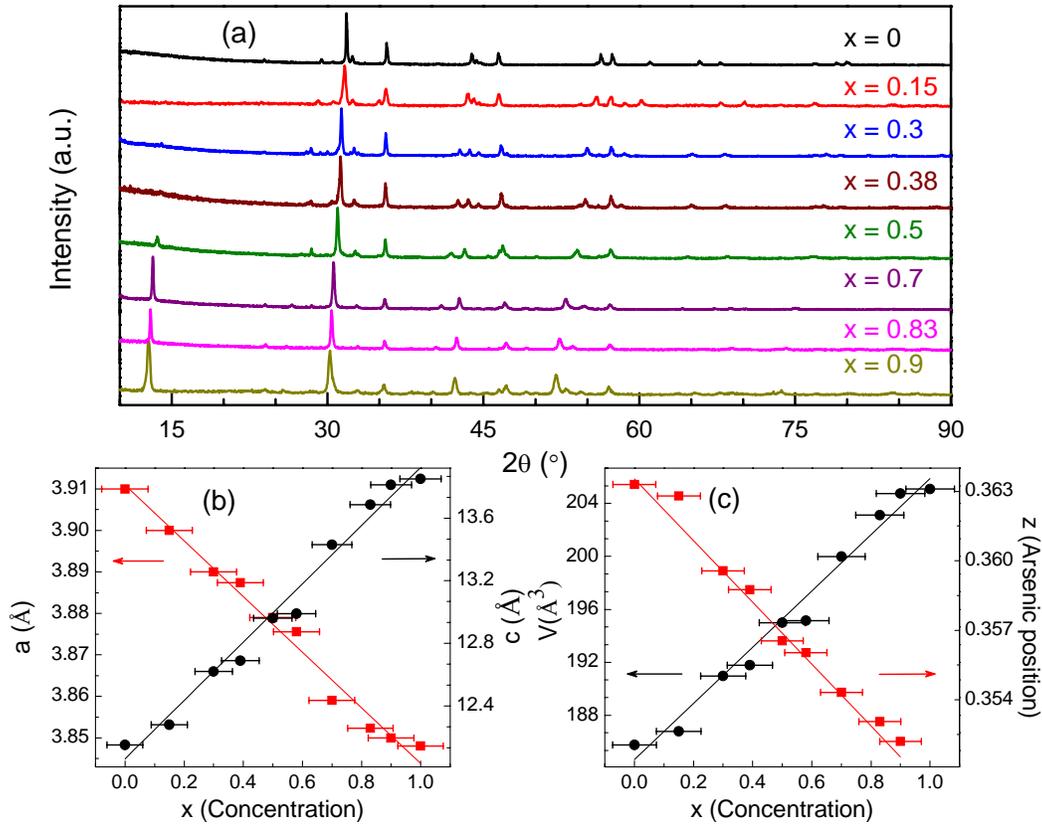

FIG. 1: (Color online) (a) X-ray powder diffraction pattern of $Eu_{1-x}K_xFe_2As_2$ (x = 0, 0.15, 0.3, 0.38, 0.5, 0.7, 0.83, 0.9) samples at room temperature. (b) and (c) represents the refined lattice parameters, unit cell volume and position of arsenic atom, z plotted as function of K content x.

**Results and Discussion**

Figure 1a shows the x-ray diffraction pattern of $Eu_{1-x}K_xFe_2As_2$ (x = 0, 0.15, 0.3, 0.38, 0.5, 0.7, 0.83, 0.9) samples taken at room temperature. The XRD pattern of all the $Eu_{1-x}K_xFe_2As_2$ samples were refined with the tetragonal $ThCr_2Si_2$ structure by least squares Rietveld refinement method using FULLPROF software. Fig.1b and 1c shows the plot of lattice parameters (a and c), unit cell volume (V) and z (position of arsenic atom) as a function of x as obtained from the Rietveld refined data. The unit cell volume increases linearly while the z value decreases with increase in x. The values of lattice parameters for pure $KFe_2As_2$ are taken from the literature [10]. Minority phase(s) detected in our samples from the un-indexed XRD peaks and SEM is less than 5 %. Compositional ratios and homogeneity of the samples were confirmed by EDAX.

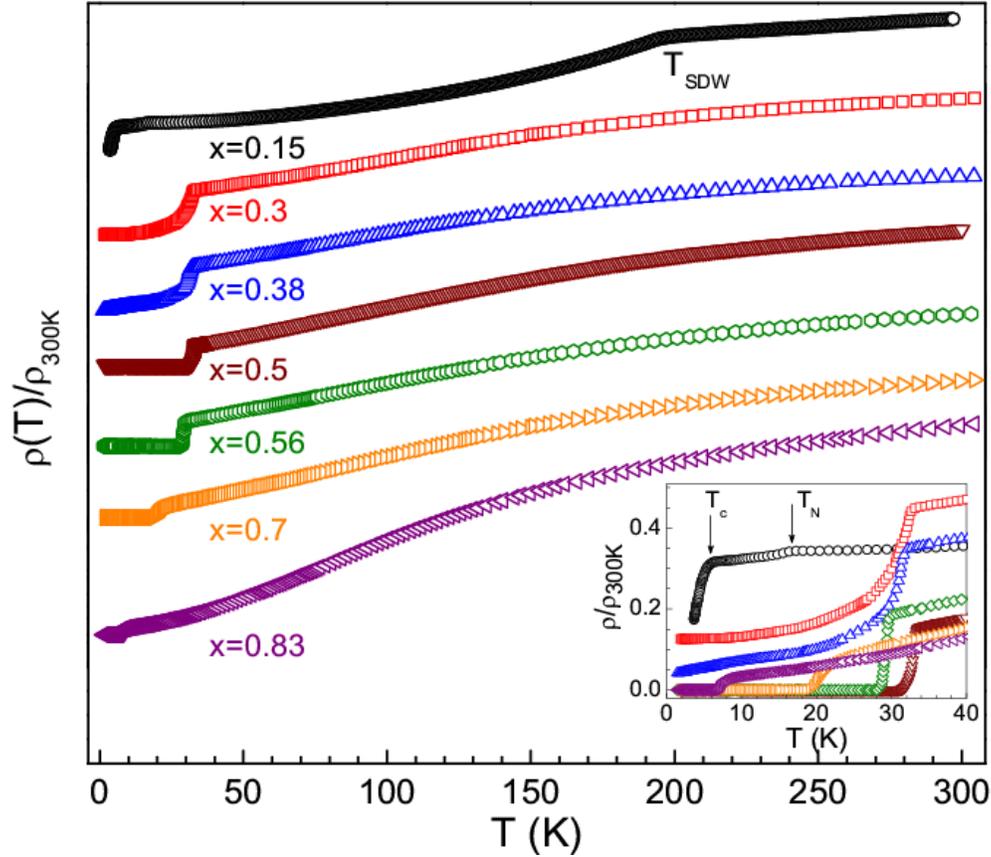

FIG. 2: (Color online) Temperature dependence of normalized electrical resistivity for the $Eu_{1-x}K_xFe_2As_2$ samples. The inset shows the expanded version of resistivity in the low temperature range (2 - 35 K).

Figure 2 shows the temperature dependence of electrical resistivity normalized to the value at room temperature, $\rho/\rho_{300K}(T)$, for the series $Eu_{1-x}K_xFe_2As_2$ (x = 0.15, 0.3, 0.38, 0.5, 0.56, 0.7, 0.8). The Curves are shifted vertically for clarity. Resistivity of pure $EuFe_2As_2$ exhibits two transitions corresponding to the SDW instability and the ordering of $Eu^{2+}$ moments, respectively, which is in good agreement with literature [5]. We find that for the lowest doping concentration x = 0.15, the signature of the SDW transition is still present and superconductivity appears at low temperature. We observe a shift in the antiferromagnetic transition to ≈ 16 K followed by a drop in the resistivity curve at 5.5 K associated with the superconducting transition. Thus the superconductivity coexists with the antiferromagnetically ordered state i.e. $T_c < T_N$. Similar behaviour has been observed in $DyNi_2B_2C$ where the onset of superconducting transition $T_c$ (~ 6.2 K) occurs below the long range antiferromagnetically ordered state of Dy moments ($T_N$ =

10.3 K) [16]. The SDW transition gets completely suppressed for x = 0.3 and resistivity decreases linearly with temperature and finally shows a drop below 32 K but zero resistance could not be reached down to 2 K. For x = 0.38 samples, we observe a slight saturation in resistivity near to 13 K which further decreases below 6 K. This might be due to the interplay of the magnetic ordering of $Eu^{2+}$ moments below 15 K which hinders the superconductivity and hence the zero resistance state. A sharp drop in resistivity is observed for x ≥ 0.5. A maximum $T_c$ of 33 K is observed for x = 0.5 which further decreases with increase in x. The inset of Fig. 2 shows the resistivity of the samples in the low temperature region (2 - 35 K).

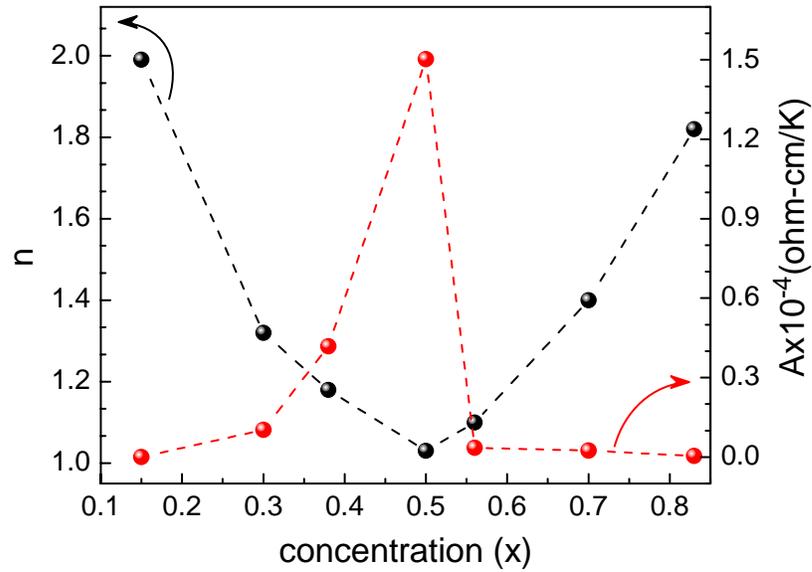

FIG. 3: (Color online) The resistivity exponent n and the coefficient A as a function of concentration x.

We fitted the low temperature part of the resistivity curves above $T_c$ up to 100 K to a power law $\rho(T) = \rho_0 + AT^n$ and the corresponding fitted parameters A and n are plotted in Fig. 3. Initially, the exponent n decreases with increase in x and approaches a value of n = 1 for $x_c$ = 0.5. For x > 0.5 the value of the exponent start to increase and approaches a value of n = 2. Thus, we see a clear crossover from Fermi to non Fermi liquid behaviour as we go towards the critical doping $x_c$ = 0.5. Similar crossover and linear T-dependence of resistivity near the critical doping level was also observed in K-doped $SrFe_2As_2$ and $BaFe_2As_2$ systems [17]. The deviation from the Fermi liquid behaviour has been observed in the transport properties of various heavy fermion systems in the regime of quantum criticality and was proposed to be due to the dominance of spin fluctuations in the electrical transport [18].

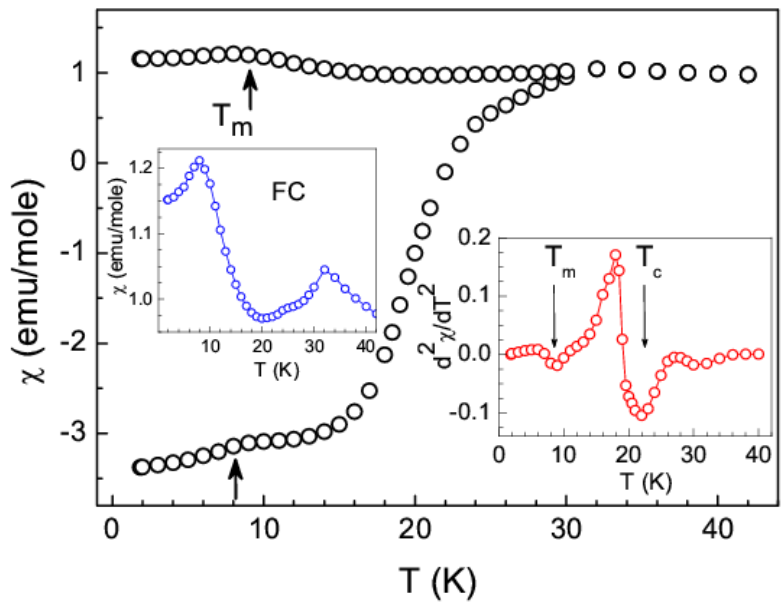

FIG. 4: (Color online) Low temperature ZFC and FC dc magnetic susceptibility for $Eu_{0.62}K_{0.38}Fe_2As_2$ under an applied field of 0.01T. The left inset shows the expanded version of FC data, while the right inset demonstrates the double derivative of ZFC magnetic susceptibility. The superconducting transition temperature $T_c$ and Eu magnetic ordering temperature $T_m$ are marked by the arrows.

To confirm the superconductivity in our samples we did the low field dc magnetization below 45 K. In Fig. 4, we show the temperature dependence of zero filed cooled (ZFC) and field cooled (FC) dc magnetic susceptibility for $Eu_{0.62}K_{0.38}Fe_2As_2$ under an applied field of 0.01 T. A clear diamagnetic signal followed by a small hump and then an increase in diamagnetic signal is observed in ZFC data. The small hump near 10 K could be due to the magnetic ordering of $Eu^{2+}$ moments which is consistent with the aforementioned reason for non-zero resistance in the superconducting state of the same sample. Under the FC condition, the magnetic susceptibility initially shows a decrease below 28 K with decrease in temperature due to the expulsion of the magnetic flux in superconducting state and then an upturn around 20 K followed by a peak at 10 K, below which it decreases again (shown in the left inset of Fig. 4). Such reentrant behaviour has also been observed in ZFC and FC measurements of $HoNi_2B_2C$ [16]. The increase of FC magnetic susceptibility below 20 K could be due to the interplay of Eu magnetic ordering. The right inset of Fig. 4 shows the plot of double derivative of ZFC magnetic susceptibility. The superconducting transition temperature $T_c$ and the magnetic ordering temperature of $Eu^{2+}$ moments $T_m$ are marked by arrows. We have also studied the field dependence of dc magnetic susceptibility under various applied field up to 0.5 T. It is found that the superconducting diamagnetic signal get weakened and finally disappeared while the peak at 10 K gets more pronounced and shifts towards lower temperature with increase in magnetic field ($T_m$ = 4 K at 0.5 T) which suggests that this peak could be due to the antiferromagnetic ordering of $Eu^{2+}$ moments. Thus, the dilution of magnetic $Eu^{2+}$ ions destabilized the antiferromagnetically ordered state.

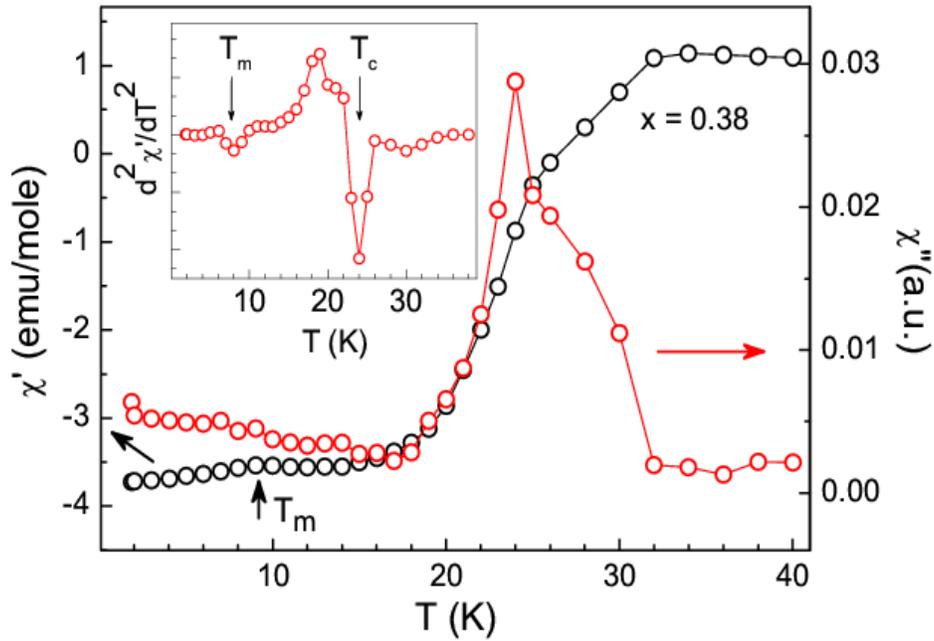

FIG. 5: (Color online) Low temperature ac magnetic susceptibility for $Eu_{0.62}K_{0.38}Fe_2As_2$. The inset depicts the double derivative of real part of ac magnetic susceptibility.

The superconductivity in K-doped $EuFe_2As_2$ is also confirmed by ac magnetic susceptibility measurements with $H_{ac}$ = 1 Oe and f = 115 Hz. Fig. 5 shows the real and imaginary parts of ac susceptibility for $Eu_{0.62}K_{0.38}Fe_2As_2$ in a temperature range 2 to 40 K. The inset of Fig. 5 shows the double derivative of real part $\chi'$. The two aforementioned peaks at $T_c$ and $T_m$ are clearly visible and are shown by arrows. The increase of imaginary part $\chi''$ below 15 K could be due to the presence of the Eu magnetic component. The T-x phase diagram obtained from the magnetic susceptibility and resistivity data is plotted in Fig. 6. A dome like parabolic $T_c(x)$ curve similar to $Ba_{1-x}K_xFe_2As_2$ and $Sr_{1-x}K_xFe_2As_2$ is observed with a tail like extension to the large value of x [9, 10]. The value of $T_c$ for $KFe_2As_2$ is obtained from the literature [10]. The highest $T_c = 33$ K is observed for K content, x = 0.5. We found a good agreement of superconducting transition temperature $T_c$ obtained from the magnetic susceptibility and resistivity curves. The coexistence of Eu moments magnetic ordering with superconductivity is another distinct feature which is absent in the Sr and Ba systems. $T_m$ decreases with increase in x and finally disappears near x = 0.6. We have also seen the coexistence of SDW and superconductivity in under doped sample viz. x = 0.15. SDW anomaly suppressed completely for x ≥0.3.

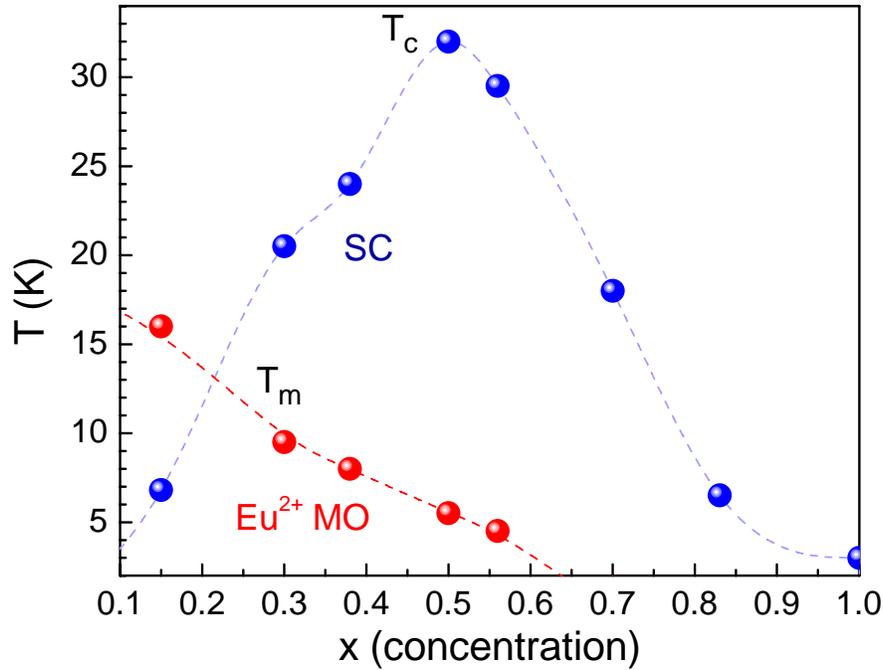

FIG. 6: (Color online) T-x phase diagram of $Eu_{1-x}K_xFe_2As_2$ with the superconducting ($T_c$) and Eu magnetic ordering ($T_m$) temperatures. $T_c$ for $KFe_2As_2$ is obtained from the literature.

Figure 7 shows the typical isothermal magnetization loops M(H) for $Eu_{1-x}K_xFe_2As_2$ (x = 0, 0.3, 0.38, 0.5) measured at 2 K. The M(H) loops are composed of two contributions: one due to the superconductivity (diamagnetic) and other due to paramagnetic component of $Eu^{2+}$ moments. The saturation magnetization is close to $7\mu_B$/Eu for all the samples which indicates the 2+ valence state of Eu ions. Thus the doping of Eu ions by non-magnetic potassium lowers the magnetic ordering temperature, without affecting the Eu valence state. The inset of Fig. 7 shows the M(H) curve for x = 0.5 sample at low fields and the deviation of the data from the linear Meissner line which is further employed to extract the lower critical field as discussed below.

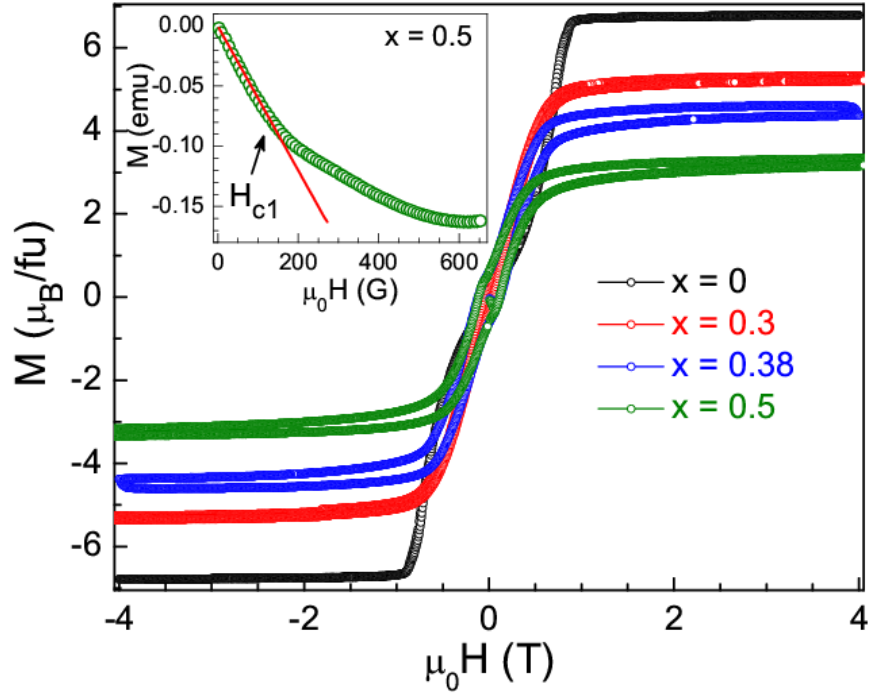

FIG. 7: (Color online) Magnetization hysteresis loops M (H) of $Eu_{1-x}K_xFe_2As_2$ (x = 0, 0.3, 0.38, 0.5) at 2 K. The inset shows the field dependence of initial magnetization curve for x = 0.5 at low fields. The solid line gives the linear fit to low field M(H) curve.

The temperature dependence of superconducting lower critical field $H_{c1}$ can be determined using low field magnetic hysteresis measurements. The deviation point of the M (H) virgin curve from the Meissner line gives the value of the first penetration field $H_{c1}^*$. We have determined the lower critical field $H_{c1}^*$ for different K doped samples viz. x = 0.3, 0.5 and 0.7. Fig. 8 displays the typical M (H) curves for x = 0.7 at selected low temperatures at low fields. It is clear that all curves at low field strength exhibits the common linear dependence with field. The fitted linear line as shown in Fig. 8 describes the Meissner shielding effects at low fields. We subtracted this Meissner line from the magnetization M for each isotherm to determine the value of $H_{c1}^*$. The inset of Fig. 8 shows the plot of Δm vs. H at different temperatures. The values of $H_{c1}^*$ at different temperatures were determined by the criterion Δm = $1×10^{-3}$ emu (resolution limit) as shown by the red line in the inset of Fig. 8. The value of lower critical field $H_{c1}$ can be derived from $H_{c1}^*$, by taking the demagnetization factor into account. For an applied external field $H_a$, the internal field $H_i = H_a - NM$ and hence $H_{c1} = H_{c1}^*/(1-N)$, where N is the demagnetization factor.

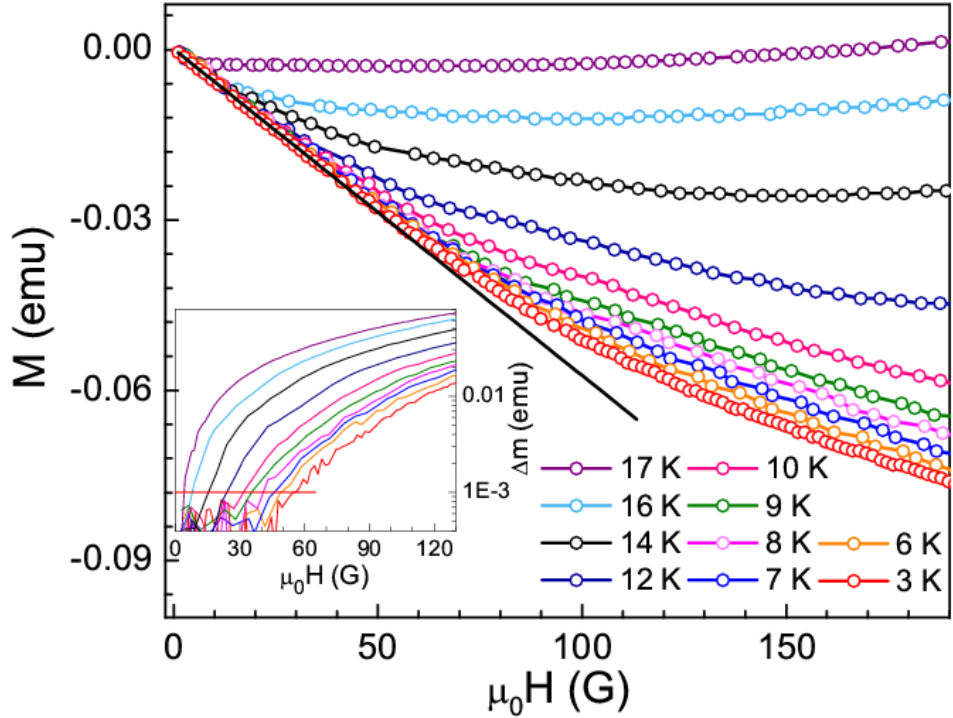

FIG. 8: (Color online) The initial part of the Magnetization curves M(H) of $Eu_{0.3}K_{0.7}Fe_2As_2$ at various temperatures. The solid line gives the Meissner linear approach. The inset shows the magnetization data subtracted by the Meissner line. The solid red line in the inset shows a criterion of $\Delta m = 1\times10^{-3}$ emu (resolution limit) used to determine $H_{c1}$.

The value of $H_{c1}$ can also be calculated using the relation between the first penetration field $H_{c1}^*$ and lower critical field $H_{c1}$ as given by Brandt [19]. It was shown that for a rectangular shaped sample lower critical field, $H_{c1} = H_{c1}^*/\tanh\left(\sqrt{0.36(b/a)}\right)$, where a and b are the width and thickness of the sample, respectively. Using this formula, we estimated the effective demagnetization factor for the above mentioned three doped samples. The value of $N_{eff}$ comes out to be 0.69, 0.68 and 0.76 for x = 0.3, 0.5 and 0.7 with samples cross-section 2.42×0.67, 2.25×0.67 and 3.01×0.49 mm$^2$, respectively. The temperature dependence of $H_{c1}$ for the three samples after taking the demagnetization factor into account is shown in Fig. 9(a). It is observed that the values of $H_{c1}$ obtained for x = 0.3 and x = 0.7 samples are nearly similar. $H_{c1}(T)$ curves decreases continuously down to the critical temperature $T_c$. The extrapolation of the $H_{c1}(T)$ data down to T = 0 K yields the value of $H_{c1}(0)$ = 248, 385 and 250 Oe for x = 0.3, 0.5 and 0.7, respectively.

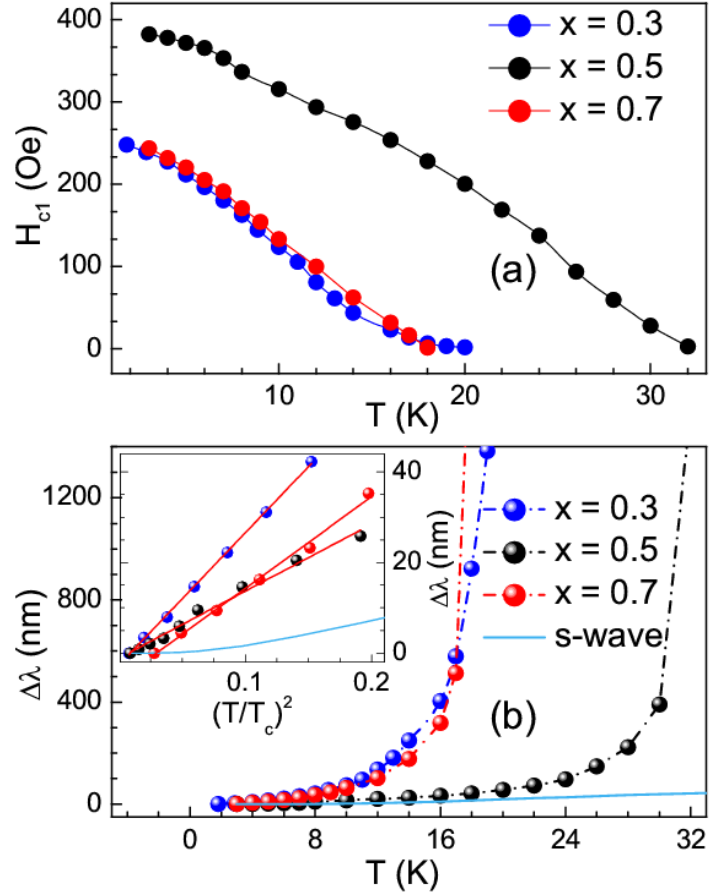

FIG. 9: (Color online) (a) The extracted $H_{c1}$ as a function of temperature for $Eu_{1-x}K_xFe_2As_2$ (x = 0.3, 0.5, 0.7) after taking the demagnetization factor into account. (b) The temperature dependence of $\Delta\lambda = \lambda(T)-\lambda(0)$ for $Eu_{1-x}K_xFe_2As_2$ calculated using the data of $H_{c1}(T)$ as described in the text. The inset shows the plot of $\Delta\lambda$ as a function of $(T/T_c)^2$ and the solid line is the calculated $\Delta\lambda$ using the s-wave isotropic BCS model.

We estimated the value of penetration depth at low temperatures using the London formula $H_{c1} = \Phi_0/(4\pi\lambda^2)(\ln\kappa+0.5)$, where $\Phi_0$ = 20.7Oe µm$^2$ is the flux quantum and $\kappa$ (= $\lambda(0)/\xi(0)$) is the Ginzburg-Landau parameter [13]. The value of $\kappa$ was determined from the equation $\dfrac{2H_{c1}(0)}{H_{c2}(0)} = \dfrac{\ln\kappa+0.5}{\kappa^2}$. Solving this equation numerically for $Eu_{0.5}K_{0.5}Fe_2As_2$ using the value of $H_{c1}(0) = 0.0385$ T and $H_{c2}(0) = 101$ T [7], we obtained $\kappa = 80$. Using this value of $\kappa$, we obtained $\lambda(0) = 180$, 145 and 179 nm for x = 0.3, 0.5 and 0.7, respectively. These values are in close agreement with the value of $\lambda(0)$ reported for $Ba_{0.6}K_{0.4}Fe_2As_2$ ($\lambda(0) = 110$ nm), Sm-1111 ($\lambda(0) = 190$ nm) and $Ba(Fe_{0.93}C_{0.07})_2As_2$ ($\lambda(0) = 208$ nm) etc [20, 21, 13]. The symmetry of superconducting pairing state is usually determined by the low temperature behaviour of the deviation in penetration depth $\Delta\lambda = \lambda(T)-\lambda(0)$ and the superfluid density, $\rho_s(T) = H_{c1}(T)/H_{c1}(0)$. In Fig. 9b we plotted the deviation in penetration depth $\Delta\lambda$ as a function of temperature for $Eu_{1-x}K_xFe_2As_2$ (x = 0.3, 0.5, and 0.7) and the curve calculated for s-wave. The inset of Fig. 9b describes the plot of $\Delta\lambda$ as a function of $(T/T_c)^2$. The red solid lines are the linear fit to the curves and hence suggest $T^2$ dependence of $\Delta\lambda$ for all the three samples at low temperatures. The quadratic temperature dependence of the penetration depth has also been observed for $Ba_{1-x}K_xFe_2As_2$ and Co doped $BaFe_2As_2$ at low temperatures [12, 13]. The $T^2$ dependence of $\Delta\lambda$ contradicts the result of an isotropic s-wave conventional superconductor for which

$$\Delta\lambda(T) = \lambda(0)\sqrt{\frac{\pi\Delta(0)}{2K_BT}} \exp\left(-\Delta(0)/K_BT\right),$$

where $\Delta(0) = 1.76\, K_BT_c$ is the maximum gap value. The penetration depth $\Delta\lambda(T)$ obtained for our samples are clearly non-exponential as shown in the inset of Fig. 9b. Also, for a pure d-wave superconductor, or any other unconventional superconductor with nodes in the gap, the penetration depth varies linearly with temperature, i.e., $\Delta\lambda(T) \sim T$. The linear T-dependence of $\Delta\lambda(T)$ has been observed for high quality single crystals of YBCO, and BSCCO-2212 etc [22, 23]. However large impurity scattering rate can change the temperature dependence of $\Delta\lambda(T)$ from T to $T^2$ due to an additional quasi-particle density of states [24]. Other factor which could lead to $\Delta\lambda(T) \sim T^2$ dependence even in the clean limit is the non-locality. The nonlocal effects dominate at low temperatures leading to divergence of effective coherence length near the nodes and hence the deviation of $\Delta\lambda(T)$ from the linear behaviour [25]. The crossover temperature below which the nonlocal effects occurs is given by $T^* = (\xi_0/\lambda_0)\Delta_0$, where $\xi_0$ is the coherence length. The nonlocal temperature $T^*$ is estimated to be $\approx$ 1- 2 K for our $Eu_{0.5}K_{0.5}Fe_2As_2$ sample with $T_c = 32$ K and $\xi_0 = 18$ Å [7]. However the quadratic temperature dependence of $\Delta\lambda(T)$ is observed up to 12 K (see inset of Fig. 9b). Thus the nonlocal effects can be completely neglected and hence the impurity scattering could be the main mechanism responsible for the observed $T^2$ dependence of $\Delta\lambda(T)$ at low temperatures.

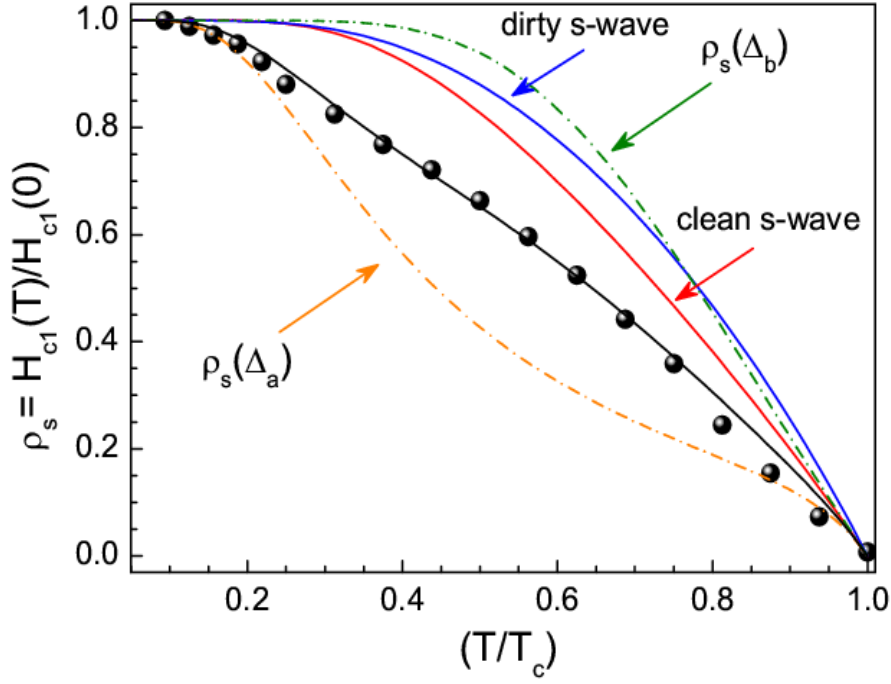

FIG. 10: (Color online) Superfluid density, $\rho_s$, as a function of normalized temperature for $Eu_{0.5}K_{0.5}Fe_2As_2$. The calculated curves for clean and dirty s-wave are shown for comparison. The solid line through the data points is the fitting curve using the two gap model as described in the text. The contributions of the small gap $\rho_s(\Delta_a)$ and the large gap $\rho_s(\Delta_b)$ in the model are also shown by the dash lines.

Figure 10 shows the superfluid density $\rho_s(T) = \lambda^2(0)/\lambda^2(T)$ for $Eu_{0.5}K_{0.5}Fe_2As_2$ along with the clean and dirty s-wave behaviour. The data clearly exhibits non exponential behaviour at low temperatures. We fit the superfluid density to a phenomenological two-gap model which has been employed in $Ba_xK_{1-x}Fe_2As_2$

$$\rho_s(T) = x\rho_s(\Delta_a) + (1-x)\rho_s(\Delta_b) \qquad (1)$$

where x is the relative weight factor for the first gap, $\Delta_a$ [26, 14]. For a single gap conventional superconductor the superfluid density $\rho_s(T)$ is given by

$$\rho_s(T) = 1 + 2\int \frac{df(E)}{dE}\left(\frac{E}{\sqrt{E^2 - \Delta(T)^2}}\right)dE \qquad (2)$$

where f(E) is the Fermi function and the total energy $E = \sqrt{\varepsilon^2 + \Delta^2}$, and $\varepsilon$ is the single particle energy measured from the Fermi surface. The continuous solid line through the data points shows the fit to equation (1) considering two independent s-wave gaps $\Delta_a$ & $\Delta_b$ using equation (2) along with their relative contributions x and 1-x, respectively. The corresponding fitting parameters are $\Delta_a$ = 2.4 meV, $\Delta_b$ = 7.72 meV, and x = 0.56. These values of gaps are almost similar to that reported for $Ba_{0.6}K_{0.4}Fe_2As_2$ ($\Delta_a$ = 2.2 meV, and $\Delta_b$ = 8.8 meV) [20]. Although the fitted two gap model and obtained parameters indicate a multi-gap nature of superconductivity in $Eu_{0.5}K_{0.5}Fe_2As_2$, the non-exponential behaviour of superfluid density and a power law behaviour for $\lambda(T)$ at low temperatures as described above requires an unconventional order parameter. The temperature dependence of the superfluid density also depends on the quality of samples as observed in $Ba_xK_{1-x}Fe_2As_2$ single crystals [14]. It is observed that the superfluid density shows exponential behaviour consistent with fully opened two gaps for the cleanest sample while the samples with large scattering rate leads to strong temperature dependence with power law behaviour at low temperatures.

**Conclusion**

In conclusion, we have synthesized a series of layered $Eu_{1-x}K_xFe_2As_2$ (x = 0-1) compounds by solid state reaction and investigated the doping dependence of magnetic and transport properties. We have seen a clear signature of the coexistence of superconducting transition ($T_c$ = 5.5 K) with SDW ordering associated with Fe-sublattice in under doped sample viz. x = 0.15. The Spin density wave anomaly is completely suppressed for x = 0.3 and onset of superconductivity appears below 20 K. A dome shaped T-x phase diagram of superconducting state is observed with a maximum $T_c$ = 33 K for x = 0.5. The coexistence of $Eu^{2+}$ magnetic ordering with superconductivity is observed up to x = 0.6. To explore the exact nature of magnetic ordering further experiments like magnetic x-ray scattering, NMR are required. The crossover from Fermi to non-Fermi liquid behaviour as seen in the transport properties near the critical doping $x_c$ = 0.5 points towards the existence of a magnetic quantum critical point. We have deduced the temperature dependence of lower critical field $H_{c1}$ from the magnetization measurements at low fields. The superfluid density for $Eu_{0.5}K_{0.5}Fe_2As_2$ can be described by s-wave two-gap model and the obtained values of gaps are $\Delta_a$ = 2.4 meV and $\Delta_b$ = 7.72 meV, respectively. The London penetration depth deviates from the exponential behaviour at low temperatures and exhibits $T^2$ dependence up to $0.4T_c$. These results can be explained either with the existence of gapless regions or point nodes or strong impurity scattering or $s^{\pm}$ pairing on the Fermi surface.